
\input lanlmac


\def\O{{\cal O}}

\def\th{\theta}
\def\Th{\Theta}
\def\cob{\delta}
\def\ep{\epsilon}

\def\S{{\bf S}}
\def\Z{{\bf Z}}
\def\tr{{\rm tr}}
\def\Tr{{\rm Tr}}

\def\hf{{1\over 2}}

\def\o{\over}

\def\si{\sigma}
\def\Si{\Sigma}
\def\b#1{\overline{#1}}
\def\del{\partial}
\def\wg{\wedge}
\def\lap{\Delta}
\def\bra{\langle}
\def\ket{\rangle}
\def\lf{\left}
\def\ri{\right}
\def\riya{\rightarrow}

\def\la{\lambda}

\def\h#1{\widehat{#1}}

\def\bt{\beta}
\def\ga{\gamma}
\def\Ga{\Gamma}
\def\al{\alpha}
\def\om{\omega}

\def\rt#1{\sqrt{#1}}

\def\st{\star}

\def\sitarel#1#2{\mathrel{\mathop{\kern0pt #1}\limits_{#2}}}

\def\Det{{\rm Det}}

\def\Pf{{\rm Pf}}

\def\np#1#2#3{{ Nucl. Phys.} {\bf B#1} (#2) #3}

\def\pln#1#2#3{{Phys. Lett.} {\bf B#1} (#2) #3}

\def\pr#1#2#3{{ Phys. Rev.} {\bf D#1} (#2) #3}

\def\hpt#1{{\tt hep-th/#1}}


\lref\cubicSFT{
E. Witten,
``Noncommutative Geometry And String Field Theory,''
Nucl.\ Phys.\  {\bf B268} (1986) 253.
}

\lref\KT{
T. Kawano and T. Takahashi,
``Open String Field Theory on Noncommutative Space,'',
\hpt{9912274};
``Open-Closed String Field Theory in the Background B-Field,''
\hpt{0005080}. 
}
\lref\Sugino{
F. Sugino,
``Witten's Open String Field Theory in Constant B-Field Background,''
JHEP {\bf 0003} (2000) 017, \hpt{9912254}. 
}

\lref\chuho{
C.-S. Chu and P.-M. Ho,
``Noncommutative Open String and D-brane,''
\np{550}{1999}{151}, \hpt{9812219};
``Constrained Quantization of Open String in Background B Field 
and Noncommutative D-brane,'' 
Nucl. Phys. {\bf B568} (2000) 447, \hpt{9906192}.
}
\lref\jab{
F. Ardalan, H. Arfaei and M. M. Sheikh-Jabbari,
`` Noncommutative Geometry From Strings and Branes,''
JHEP {\bf 9902} (1999) 016, \hpt{9810072};
``Dirac Quantization of Open Strings and Noncommutativity in Branes,''
Nucl. Phys. {\bf B576} (2000) 578, \hpt{9906161}\semi
M. M. Sheikh-Jabbari,
``Open Strings in a B-field Background as Electric Dipoles,''
\pln{455}{1999}{129}, \hpt{9901080}.
}
\lref\scho{
V. Schomerus,
`` D-branes and Deformation Quantization,''
JHEP {\bf 9906} (1999) 030, \hpt{9903205}.
}
\lref\ACNY{
A. Abouelsaood, C. G. Callan, C. R. Nappi and S. A. Yost,
``Open Strings in Background Gauge Fields,''
\np{280}{1987}{599}.
}

\lref\Ishio{
N. Ishibashi,
``$p$-branes from $(p-2)$-branes in the Bosonic String Theory,''
\np{539}{1999}{107}, \hpt{9804163}.
}
\lref\branes{
T. Banks, N. Seiberg and S. Shenker,
``Branes from Matrices,''
\np{490}{1997}{91}, \hpt{9612157}.
}
\lref\BFFS{
T. Banks, W. Fischler, S. H. Shenker and L. Susskind,
``M Theory As A Matrix Model: A Conjecture,''
\pr{55}{1997}{5112}, \hpt{9610043}.
}
\lref\town{
P. K. Townsend,
``D-branes from M-branes,''
\pln{373}{1996}{68}, \hpt{9512062}.
}

\lref\SW{
N. Seiberg and E. Witten,
``String Theory and Noncommutative Geometry'',
JHEP {\bf 9909} (1999) 032, \hpt{9908142}.
}

\lref\Witten{
E. Witten,
``On Background Independent Open-String Field Theory,''
Phys. Rev. {\bf D46} (1992) 5467, \hpt{9208027}.
}
\lref\WittenT{
E. Witten,
``Some Computations in Background Independent Open-String Field Theory,''
Phys. Rev. {\bf D47} (1993) 3405, \hpt{9210065}.
}

\lref\Shata{
S. L. Shatashvili,
``Comment on the Background Independent Open String Theory,''
Phys. Lett. {\bf B311} (1993) 83, \hpt{9303143};
``On the Problems with Background Independence in String Theory,''
\hpt{9311177}.
}
\lref\LW{
K. Li and E. Witten,
``Role of Short Distance Behavior in Off-Shell Open-String Field Theory,''
Phys. Rev. {\bf D48} (1993) 853, \hpt{9303067}.
}

\lref\KutasovMM{
D. Kutasov, M. Marino and G. Moore,
``Some Exact Results on Tachyon Condensation in String Field Theory,''
\hpt{0009148}.
}
\lref\GerasimovS{
A. A. Gerasimov and S. L. Shatashvili,
``On Exact Tachyon Potential in Open String Field Theory,''
\hpt{0009103}.
}
\lref\GhoshalSen{
D. Ghoshal and A. Sen,
``Normalization of the Background Independent Open String Field 
Theory Action,''
\hpt{0009191}.
}
\lref\CornalbaSFT{
L. Cornalba,
``Tachyon Condensation in Large Magnetic Fields with Background Independent
String Field Theory,''
\hpt{0010021}.
}

\lref\GMS{
R. Gopakumar, S. Minwalla and A. Strominger,
``Noncommutative Solitons,''
JHEP {\bf 0005} (2000) 020, \hpt{0003160}.
}
\lref\WittenNCG{
E. Witten,
``Noncommutative Tachyons And String Field Theory,''
\hpt{0006071}.
}
\lref\Dasgupta{
K. Dasgupta, S. Mukhi and G. Rajesh,
``Noncommutative Tachyons,''
JHEP {\bf 0006} (2000) 022, \hpt{0005006}.
}
\lref\Harvey{
J. A. Harvey, P. Kraus, F. Larsen and E. J. Martinec,
``D-branes and Strings as Non-commutative Solitons,''
JHEP {\bf 0007} (2000) 042, \hpt{0005031}.
}

\lref\GMStwo{
R. Gopakumar, S. Minwalla and A. Strominger,
``Symmetry Restoration and Tachyon Condensation in Open String Theory,''
\hpt{0007226}.
}
\lref\SenNothing{
``Some Issues in Non-commutative Tachyon Condensation,''
\hpt{0009038}.
}

\lref\Tseytlin{
A. A. Tseytlin,
``Born-Infeld Action, Supersymmetry and String Theory,''
\hpt{9908105}.
}
\lref\FradTsey{
E. S. Fradkin and A. A. Tseytlin,
``Quantum String Theory Effective Action,''
Nucl.\ Phys.\  {\bf B261} (1985) 1.
}

\lref\SeiMat{
N. Seiberg,
``A Note on Background Independence in Noncommutative Gauge Theories, Matrix
Model and Tachyon Condensation,''
JHEP {\bf 0009} (2000) 003, \hpt{0008013}.
}

\Title{             
                                             \vbox{\hbox{KEK-TH-719}
                                             \hbox{hep-th/0010028}}}
{\vbox{
\centerline{Noncommutative Tachyon from} 
\vskip 4mm
\centerline{Background Independent Open String
Field Theory}
}}

\vskip .2in

\centerline{
                    Kazumi Okuyama
}

\vskip .2in

\centerline{\sl Theory Group, KEK, Tsukuba, Ibaraki 305-0801, Japan}
\centerline{\tt kazumi@post.kek.jp }

\vskip 3cm
\noindent

We analyze the tachyon field in 
the bosonic open string theory in a constant
$B$-field background using the background independent open string 
field theory. 
We show that in the large noncommutativity limit the action of tachyon field  
is given exactly by the potential term which has the same form as
in the case without $B$-field but the product of tachyon field 
is taken to be the star product.

\Date{October 2000}

\vfill
\vfill

\newsec{Introduction}

The problem of describing the process of tachyon condensation
has attracted many people. By turning on a constant $B$-field, 
one can handle the behavior of tachyon by using  
the so-called noncommutative tachyon \refs{\GMS,\Dasgupta,\Harvey}
and it enables us to  construct the lower dimensional brane easily as a 
topological defect on an unstable brane containing tachyonic modes. 
To understand the fate of tachyon more thoroughly we need the information
of the form of tachyon potential.
Very recently, the problem of the tachyon potential
was studied in \refs{\GerasimovS,\KutasovMM,\GhoshalSen} 
using the background independent 
open string field theory \refs{\Witten,\WittenT,\Shata}.

In this paper, combining these two ideas we consider tachyon field
in the bosonic  open string theory, or tachyon on the $D25$-brane,
in a constant $B$-field background and calculate the action of tachyon
field using the background independent open string field theory. 
We show that in the large noncommutativity limit the potential of the form
$ e^{-T}(T+1)$ with the product of fields taken by the star product 
gives the exact action of tachyon.

This paper is organized as follows: In section 2, we calculate the 
action of tachyon field in a $B$-field background following \WittenT.
In section 3, we study this action in the derivative expansion and 
in the large noncommutativity limit.
Section 4 is devoted to discussions.

\newsec{Open String Field Theory in a Constant $B$-Field Background}
In this section, we calculate the action of tachyon field in the presence of 
a background $B$-field using the background independent open string field 
theory following 
the procedure in \WittenT. As we will see,  
most of the calculation are parallel
to those in \WittenT\ without $B$-field. 

\subsec{Green's Function}
The worldsheet action in the bulk of the disc $\Si$ in the background of 
the metric $g_{\mu\nu}$ and the $B$-field $B_{\mu\nu}$ is given by 
\eqn\Sbulk{
S_{\Si}={1\o4\pi\al'}\int_{\Si} d^2\xi\rt{h}h^{ab}g_{\mu\nu}\del_aX^{\mu}
\del_bX^{\nu}
-{i\o2}\int_{\Si}B_{\mu\nu}dX^{\mu}\wg dX^{\nu}.
}
We take $\Si$ to be a unit disc $\{|z|\leq 1\}$  and the worldsheet metric 
to be flat:
\eqn\flath{
h_{ab}d\xi^ad\xi^b=dzd\b{z}=dr^2+r^2d\si^2
}
where $z=re^{i\si}$. 
In the formalism of \Witten, the space of string field is identified as  
the space of boundary deformation. For the tachyon field $T(X)$,
the boundary action is given by  
\eqn\SB{
S_{\del\Si}=\int_0^{2\pi}{d\si\o2\pi}T(X(\si)).
}
Following \WittenT, we consider the tachyon configuration which is
quadratic in $X^{\mu}$: 
\eqn\tachyon{
T(X)=a+{1\o2\al'}u_{\mu\nu}X^{\mu}X^{\nu}
}
with $u_{\mu\nu}=u_{\nu\mu}$. Then the worldsheet theory is free and 
 exactly solvable.
From $S_{\Si}$ and $S_{\del\Si}$, the boundary condition
of $X^{\mu}$ becomes
\eqn\BCX{
\Big(g_{\mu\nu}\del_rX^{\nu}-i2\pi\al'B_{\mu\nu}\del_{\si}X^{\nu}
+u_{\mu\nu}X^{\nu}\Big)
\Big|_{r=1}=0.
}

To calculate the action of tachyon field, the Green's function
\eqn\propM{
M^{\mu\nu}(z,w)=\Big\bra X^{\mu}(z)X^{\nu}(w)\Big\ket
}
plays a crucial role.
$M(z,w)$ should satisfy the Laplace equation
\eqn\lapeq{
-{1\o2\pi\al'}g_{\mu\nu}\lap_zM^{\nu\rho}(z,w)=\cob^2(z-w)\cob_{\mu}^{\rho}
}
and the boundary condition \BCX.
By a straightforward calculation, we find the Green's function to be\foot{
In the case $g_{\mu\nu}=\cob_{\mu\nu}$, the Green's function in  $B$-field
background was obtained in \LW.}
\eqn\Mform{\eqalign{
{2\o\al'}M(z,w)=&\,\,
G^{-1}F_N(z,w)-\Th G\Th F_D(z,w)+2\pi i\Th F_{\ep}(z,w) \cr
&+2u^{-1}-2\sum_{k=1}^{\infty}{1\o k}\lf({E_+u\o k+E_+u}E_+(z\b{w})^k
+E_{-}{uE_{-}\o k+uE_{-}}(\b{z}w)^k\ri)
}}
where 
\eqn\defFNDE{\eqalign{
F_N(z,w)&=-\log|z-w|^2-\log|1-z\b{w}|^2,\cr
F_D(z,w)&=-\log|z-w|^2+\log|1-z\b{w}|^2, \cr
F_{\ep}(z,w)&={1\o\pi i}\log\lf({1-\b{z}w\o 1-z\b{w}}\ri),
}}
and $G,\Th$  are the open string parameters defined by 
\eqn\GTh{
{1\o g+2\pi\al'B}=G^{-1}+\Th.
}
$E_{\pm}$ in \Mform\ denotes the combination
\eqn\defEpm{
E_{\pm}=G^{-1}\pm\Th.
}
In the component notation,  $G=(G_{\mu\nu})$ and $G^{-1}=(G^{\mu\nu})$.


The Green's function at the boundary of $\Si$ becomes
\eqn\Bprop{
{2\o\al'}M(\si,\si')=2u^{-1}+2\sum_{k=1}^{\infty}\lf[
{1\o k+E_{+}u}E_{+}e^{ik(\si-\si')}+E_{-}{1\o k+uE_{-}}e^{-ik(\si-\si')}\ri].
}
Note that $F_{\ep}$ at the boundary of $\Si$ is reduced to
the sign function $\ep(\si)$ 
\eqn\signfun{
F_{\ep}(e^{i\si_1},e^{i\si_2})\riya \ep(\si_1-\si_2)
}
in the limit  $\si_1-\si_2\riya 0$.
From this relation, the boundary coordinates become noncommutative 
\refs{\chuho,\jab,\scho,\ACNY}
\eqn\xcomm{
[X^{\mu}(\si),X^{\nu}(\si)]=i\th^{\mu\nu}
}
where $\th$ is related to $\Th$ by
\eqn\smallth{
\Th^{{\mu\nu}}={\th^{{\mu\nu}}\o2\pi\al'}.
}

\subsec{Partition Sum}
The second step to calculate the action is to calculate the 
partition sum $Z$   on the disc  which is defined by
\eqn\Zsum{
Z(a,u)=\int{\cal D}X e^{-S_{\Si}-S_{\del\Si}}=e^{-a}Z(u).
}
Here we factored out the dependence of the zero mode $a$ of tachyon field. 
To calculate $Z$, we first consider the derivative of it:
\eqn\udelZ{
{d\o du_{\mu\nu}}\log Z(u)=-{1\o4\pi\al'}\int_0^{2\pi}d\si 
\Big\bra X^{\mu}(\si)X^{\nu}(\si)\Big\ket.
}
We define the Green's function at the same point by subtracting the 
divergent part
\eqn\reg{
\Big\bra X^{\mu}(\si)X^{\nu}(\si)\Big\ket=\lim_{\cob\riya 0}\lf[\Big\bra 
X^{\mu}(\si+\cob)X^{\nu}(\si)\Big\ket-{\al'\o2}C^{\mu\nu}(\cob)\ri]
}
where
\eqn\subtract{\eqalign{
C(\cob)&=G^{-1}F_N(\si+\cob,\si)+2\pi i\Th F_{\ep}(\si+\cob,\si) \cr
&=-2E_{+}\log(1-e^{i\cob})-2E_{-}\log(1-e^{-i\cob}).
}}
With this regularization we find
\eqn\Zdeluu{
{d\o du}\log Z(u)=-\hf u^{-1}+\hf\sum_{k=1}^{\infty}
{1\o k}\lf({E_{+}u\o k+E_{+}u}E_{+}+E_{-}{uE_{-}\o k+uE_{-}}\ri).
}
Using the identity of Gamma function
\eqn\Gaid{
{d\o d x}\log\Ga(x)=-{1\o x}-\ga+\sum_{k=1}^{\infty}{x\o k(k+x)},
}
\Zdeluu\  becomes
\eqn\ZdelGadel{
{d\o du}\log Z(u)=\hf u^{-1}+\hf{d\o du}\Big(\tr\log\Ga(E_{+}u)
+\tr\log\Ga(uE_{-})\Big)+\hf\ga(E_{+}+E_{-}).
}
Here $\ga$ is the Euler's constant.
We use the notation `tr' and `det' for the trace and the determinant over
the spacetime indices $\mu,\nu$.
By integrating this relation,  we find
\eqn\Ztotal{\eqalign{
Z(a,u)&=e^{-a+\ga\tr(G^{-1}u)}\det{}^{\hf}
\Big(\Ga(E_{+}u)\Ga(uE_{-})uE_{-}\Big) \cr
&=e^{-a+\ga\tr(G^{-1}u)}\det{}^{\hf}
\Big(\Ga(E_{+}u)\Ga(1+uE_{-})\Big).
}}
Here we include the factor $\det{}^{\hf}(E_{-})$ which cannot
be determined from \ZdelGadel. We will see in section 3 that this factor
is needed to reproduce the Born-Infeld action. In principle we can calculate
the normalization of $Z$ as reviewed in \Tseytlin, but we do not discuss 
it here.

Using the identity for an arbitrary function $f(x)$ 
and finite size matrices $A,B$ 
\eqn\matid{
\det \Big(f(BAB^{-1})\Big)=\det \Big(f(A)\Big), \quad 
\det\Big(f(A)\Big)=\det \Big(f(A^T)\Big),
} 
$Z$ can be written as
\eqn\Zinsimple{
Z(a,u)=e^{-a+\ga\tr(G^{-1}u)}\det\Big((E_{+}u)^{\hf}\Ga(E_{+}u)\Big).
}
Note that $Z(a,u)$ depends on the $B$-field only through
the combination $E_{\pm}$.

\subsec{Evaluation of Action}
As the final step, we construct the action of tachyon field in a
$B$-field background.
The  action $S$ of open string field theory is given by \Witten
\eqn\defrelS{
dS=\hf\int_0^{2\pi}{d\si d\si'\o (2\pi)^2}\Big\bra d\O(\si)
\,\{Q_B,\O\}(\si')\Big\ket.
}
For the generic boundary deformation
\eqn\deformV{
\int_0^{2\pi}{d\si\o2\pi}{\cal V},
}
$\O$ and ${\cal V}$ are related by $\O=c{\cal V}$. The exterior derivative
$d$ is taken on the couplings in the boundary interaction ${\cal V}$,
or $a$ and $u$ in this case.  
The BRST transformation of ${\cal O}$ corresponding to the tachyon 
is given by
\eqn\QbT{
\{Q_B,cT(X)\}=c\del_{\si}c(1-\lap_T)T(X)
}
where $\lap_T$ is the dimension of $T$. As a differential operator acting on 
$T(X)$, $\lap_T$
can be written as
\eqn\laponT{
\lap_T=-\al'G^{\mu\nu}{\del^2\o\del X^{\mu}\del X^{\nu}}.
}
Therefore
\eqn\QBTforquad{
\{Q_B,cT(X)\}=c\del_{\si}c\lf(\tr(G^{-1}u)+a
+{1\o2\al'}u_{\mu\nu}X^{\mu}X^{\nu}\ri).
}

Using the ghost correlation function
\eqn\ghostcor{
\bra c(\si)c(\si')\del_{\si'}c(\si')\ket=2\big[\cos(\si-\si')-1\big],
}
\defrelS\ becomes
\eqn\dSforau{\eqalign{
dS=&\int_0^{2\pi}{d\si d\si'\o(2\pi)^2}\big(\cos(\si-\si')-1\big) \cr
&\times\lf\bra\lf(da+{1\o2\al'}du_{\mu\nu}X^{\mu}X^{\nu}(\si)\ri)
\lf(a+\tr(G^{-1}u)+{1\o2\al'}u_{\rho\tau}X^{\rho}X^{\tau}(\si')\ri)\ri\ket.
}}

To show that the right-hand-side of \dSforau\ is an exact form,
we need an identity corresponding to eq.(2.22) in \WittenT:
\eqn\Keyid{
\int_0^{2\pi}{d\si d\si'\o(4\pi\al')^2}\cos(\si-\si')du_{\mu\nu}
u_{\rho\tau}\Big\bra
X^{\mu}X^{\nu}(\si)X^{\rho}X^{\tau}(\si')\Big\ket
=\tr(G^{-1}du).
}
Using the following relations
\eqn\cortoZ{\eqalign{
&\int_0^{2\pi}{d\si d\si'\o(2\pi)^2}\big(\cos(\si-\si')-1\big)
{1\o2\al'}\Big\bra X^{\mu}X^{\nu}(\si)\Big\ket={\del\o\del u_{\mu\nu}}Z, \cr
&\int_0^{2\pi}{d\si d\si'\o(4\pi\al')^2}
\Big\bra X^{\mu}X^{\nu}(\si)X^{\rho}X^{\tau}(\si')\Big\ket
={\del^2\o\del u_{\mu\nu}\del u_{\rho\tau}}Z,
}}
and \Keyid, we finally find that $S(a,u)$ is related to $Z(a,u)$ by
\eqn\SFTact{
S(a,u)=\lf[\tr(G^{-1}u)-a{\del\o\del a}-\tr\lf(u{\del\o\del u}\ri)+1\ri]Z(a,u).
}

\newsec{Noncommutative Tachyon}
In this section, we examine the action of tachyon field given by \SFTact.
As shown in \SW, a $D$-brane in a $B$-field background can be described
by either  commutative or noncommutative language. 
We will show that the derivative expansion of the action \SFTact\  
leads to the commutative description. On the other hand, 
by taking the large noncommutativity limit $S(a,u)$  
reproduce the action of the noncommutative tachyon. 
In the following discussion, 
we are not careful about the overall normalization of the action.
See \refs{\GhoshalSen,\KutasovMM} for recent discussions on 
the normalization of action. See also \FradTsey\ for the early discussion
on the action of tachyon in a constant $B$-field background.

\subsec{Commutative Description of Tachyon}
Let us consider the case of nearly constant tachyon, i.e.,  $u\sim 0$.
Then the expansion of $S(a,u)$ with respect to $u$ corresponds to 
the derivative expansion of tachyon field.
When $u$ is small, partition sum $Z(a,u)$ becomes 
\eqn\Zuzero{
Z(a,u)=e^{-a}\det{}^{-\hf}(E_{+}u)+\cdots = 
T_{D25}\int d^{26}x\,{\cal L}_{BI}(B)\,e^{-T(x)}
+\cdots
}
with ${\cal L}_{BI}(B)=\rt{\det(g+2\pi\al'B)}$ and $T_{D25}=(2\pi\al')^{-13}$
in this normalization of $Z$.
However, remember that the overall normalization of $Z$ 
cannot be determined within this framework.
What we can say at most is that $T_{D25}$ is proportional to $(\al')^{-13}$
from the dimensional analysis. 
The dots in \Zuzero\   denote the higher
order terms in $u$, or the higher derivative terms of $T(x)$.
(See \KutasovMM\  for the structure of the higher derivative terms.)

From this form of partition sum, we can calculate the action of tachyon field 
$S(a,u)$. 
The first term in \SFTact, which originates from $-\lap_T$, 
gives the kinetic term for the tachyon and
the other terms correspond to the potential.
First we calculate the kinetic term. Using the relation
\eqn\kinT{
\tr(G^{-1}u)=-\lap_TT=\al'G^{\mu\nu}\del_{\mu}\del_{\nu}T,
}
we find that the first term in \SFTact\  is related to the kinetic term
of $T(x)$ by 
\eqn\kintoZ{
\tr(G^{-1}u)\int d^{26}x\,e^{-T}=
\int d^{26}x\,e^{-T}\al'G^{\mu\nu}\del_{\mu}\del_{\nu}T
=\int d^{26}x\,e^{-T}\al'G^{\mu\nu}\del_{\mu}T\del_{\nu}T.
}
The differential operator in the second and the third term in \SFTact\  
generates the scale transformation $T(x)\riya \la T(x)$.
Therefore, we find
\eqn\scaledelT{
\lf[-a{\del\o\del a}-\tr\lf(u{\del\o\del u}\ri)\ri]\int d^{26}x\,e^{-T(x)}
=-\lf.{d\o d \la}\ri|_{\la=1}\int d^{26}x\,e^{-\la T(x)}
=\int d^{26}x \,T(x)e^{-T(x)}.
}
Combining \kintoZ\ and \scaledelT, the action is found to be
\eqn\Tachyaccom{
S=T_{D25}\int d^{26}x\,{\cal L}_{BI}(B)\,e^{-T}\Big(\al'G^{\mu\nu}
\del_{\mu}T\del_{\nu}T+T+1\Big)+\cdots.
}
Since the $B$-dependence in the above action  
is given by the Born-Infeld form,
this expansion corresponds to the commutative description of $D25$-brane.

\subsec{Large Noncommutativity Limit}
As was pointed out in \GMS, in the large noncommutativity limit,
the problem of the tachyon condensation is drastically simplified
since the kinetic term of tachyon disappears in this limit.
We show that this phenomenon also occurs in $S(a,u)$ and the structure
of star product emerges, which cannot be seen in the derivative expansion.

By the large noncommutativity limit, we mean  the situation
\eqn\GvsTh{
G^{-1}\ll\Th
}
or equivalently $G^{-1}$ is set to zero while $\Th$ is kept finite.  
In this limit, the partition sum $Z(a,u)$ becomes
\eqn\Zinlargeth{\eqalign{
\lim_{G\Th\riya\infty}Z(a,u)
&=e^{-a}\det{}^{\hf}\Big(\Ga(\Th u)\Ga(1-\Th u)\Big)\cr
&=e^{-a}\det{}^{\hf}\lf({\pi\o \sin\pi\Th u}\ri).
}}

To show that this form of $Z$ leads to the noncommutative description
of tachyon, let us introduce the quantity $\Xi(a,u)$ by  
\eqn\defXi{
\Xi(a,u)=\int {d^{26}x\o \Pf(2\pi\th)}\exp_{\st}(-T(x))
}
where $\exp_{\st}$ means that 
the product of $T(x)$ is taken by the star product
\eqn\defstar{
f\st g=f\exp
\lf({i\o2}\overleftarrow{\del_{\mu}}\th^{\mu\nu}
\overrightarrow{\del_{\nu}}\ri)g.
}
In the operator language, $\Xi$ is written as 
\eqn\traceXi{
\Xi(a,u)=\Tr_{{\cal H}} \exp\big(-T(\h{x})\big)
}
where the trace is taken over the Hilbert space ${\cal H}$ on which 
the operators $\h{x}^{\mu}$ satisfy the relation
\eqn\opxcom{
[\h{x}^{\mu},\h{x}^{\nu}]=i\th^{\mu\nu}.
}
Eq.\traceXi\  shows that 
$\Xi$ can be interpreted as a thermal partition function on the phase space 
$\{x^{\mu}\}$ with Hamiltonian $T(x)/2\pi$ and inverse temperature $\bt=2\pi$.
Therefore, $\Xi$ can be written as a path integral on $\S^1$:
\eqn\pathint{
\Xi(a,u)=\int{\cal D}x\exp\lf(\int_0^{2\pi}d\si{i\o2}x^{\mu}(\si)
(\th^{-1})_{\mu\nu}
\del_{\si}x^{\nu}(\si)-{T(x(\si))\o2\pi}\ri).
}
Up to a proportionality constant, $\Xi$ can be evaluated as
\eqn\Xieval{\eqalign{
\Xi(a,u)&=e^{-a}\int{\cal D}x\exp\lf(\int_0^{2\pi}d\si{i\o2}x^{\mu}(\si)
(\th^{-1})_{\mu\nu}
\del_{\si}x^{\nu}(\si)-{u_{\mu\nu}x^{\mu}(\si)x^{\nu}(\si)\o4\pi\al'}\ri) \cr
&=e^{-a}\Det^{-\hf}\lf(-i\th^{-1}\del_{\si}+{u\o2\pi\al'}\ri) \cr
&\approx e^{-a}\Det^{-\hf}(-i\del_{\si}+\Th u) \cr
&=e^{-a}\det{}^{-\hf}\prod_{n\in \Z}(n+\Th u) \cr
&\approx {e^{-a}\o\det{}^{\hf}\big(\sin\pi\Th u\big)}.
}}
The proportionality constant can be fixed from the behavior of
the right-hand-side of \defXi\ in the limit $\th\riya0$.
Then we find that $\Xi$ is given by\foot{
This form can also be deduced from the partition function of
a harmonic oscillator 
\eqn\Zosci{
Z_{osci}=\Tr e^{-\bt H}={1\o 2\sinh\lf(\hf\bt\hbar\om\ri)}
}
with 
$H=\hf (p^2+\om^2 x^2)$ 
and $[x,p]=i\hbar$.
The difference between sin and sinh in $\Xi$ and $Z_{osci}$ 
comes from the fact
that the eigenvalues of $\th$ correspond to $\pm i\hbar$.
}
\eqn\Xifinal{
\Xi(a,u)={e^{-a}\o\det{}^{\hf}\big(2\sin\pi\Th u\big)}.
}
From \Zinlargeth\ and \Xifinal, 
we conclude that in the large noncommutativity limit $Z(a,u)$ is equal to
$\Xi(a,u)$ up to a normalization factor:
\eqn\largeThZXi{
\lim_{G\Th\riya\infty}Z(a,u)=\Tr_{{\cal H}}\,\exp(-T(\h{x}))
=\int{d^{26}x\o\Pf(2\pi\th)}
\exp_{\st}(-T(x)).
}

Since $G^{-1}\ll \Th$ is equivalent to $g\ll 2\pi\al'B$,
the large noncommutativity limit can be rephrased as the large $B$ limit. 
In this limit, the bulk worldsheet action 
\Sbulk\ is given by the $B$-field term alone, which is a total derivative.
Therefore, we expect that
the partition sum $Z$ is reduced to the quantum mechanics on 
the boundary of $\Si$, which is nothing but \pathint\ 
because $\th=B^{-1}$ in this limit.
Our result \largeThZXi\ strongly suggests that   
the regularization \reg\ used in the calculation of $Z$ 
is the correct choice, since it leads to the expected result
in the large $B$ limit.

From the relation \largeThZXi, 
we can calculate the action of $T$. Since we have set $G^{-1}=0$,
the first term in \SFTact\ is zero. Using the relation
\eqn\btforT{
\lf[-a{\del\o\del a}-\tr\lf(u{\del\o\del u}\ri)\ri]\Tr_{{\cal H}}\exp(-T(\h{x}))
=\Tr_{{\cal H}}T(\h{x})\exp(-T(\h{x}))
}
the action of $T(x)$ in the large noncommutativity limit is found to be
\eqn\NCact{
S=\Tr_{{\cal H}}\Big((T(\h{x})+1)\exp(-T(\h{x}))\Big)
=\int{d^{26}x\o\Pf(2\pi\th)}(T(x)+1)\st\exp_{\st}(-T(x)).
}
As is clear from \Zinlargeth, this action contains terms 
of all order in $u$ and is exact in this limit.
Note that  the higher derivative
terms represented by dots in \Tachyaccom\  
appear only through the star product.
In \SW, it was shown 
that the S-matrix depends on $B$-field only through the Moyal phase in the
noncommutative description. Our result can be thought of as an off-shell
extension of the argument in \SW. 

In this subsection, we assumed that $\Th$ has the maximal rank.
In the lower rank case, the derivatives of tachyon vanish 
along the directions of non-zero
$\Th$  in the large $\Th$ limit.

\newsec{Discussion}
In this paper, we constructed the action of tachyon field in a $B$-field
background using the background independent open string field theory.
In the large noncommutativity limit, we found that the action of tachyon
is given by the potential term which has the same form
as in the case of vanishing $B$-field 
but the product is taken by the star product.

It will be important to compare our result to the 
cubic open string field theory \cubicSFT\  
in a constant $B$-field background studied in \refs{\KT,\Sugino}.
It may be also interesting to study
the relation to the discussion in \WittenNCG\ which says that
in the large noncommutativity limit
a string field factorizes into a oscillator part and  
an element of the algebra representing the noncommutative geometry.
Another interesting problem is the tachyon condensation in the $D\b{D}$ 
system. But due to the non-Abelian nature of Chan-Paton factors
it seems that the calculation is not so straightforward. 
In a $B$-field background, a $D$-brane can be described as a collection of
infinitely many lower dimensional branes \refs{\town,\BFFS,\branes,\Ishio}. 
The relation between this Matrix Theory 
picture and the open string field theory deserves to be  studied further.

We comment on the kinetic term of tachyon field in the case of 
non-zero $G^{-1}$. 
As is well known, the noncommutative gauge theory naturally appears
in  Matrix Theory  when it is expanded around the noncommutative 
classical solution.
In this picture, the derivative of a field is given by the commutator
of the matrix coordinate and that field, and it becomes naturally
the noncommutative covariant derivative.   
Therefore, it may be important to include the gauge field into the analysis
of tachyon field and study the background independent description using the 
matrix variables along the line of  \SeiMat.

\vskip 8mm
\noindent
{\bf Note added:} After this work was completed, we received 
a paper \CornalbaSFT\
which has some overlaps with ours.

\vskip 12mm
\centerline{{\bf Acknowledgments}}
I would like to thank N. Ishibashi  
for useful comments and discussions. 
This work was supported in part by JSPS Research Fellowships for Young
Scientists.

\listrefs

\end